\input harvmac
\input epsf
\input amssym

\def\zlambda{\lambda'\,{}}

\def\bb{
\font\tenmsb=msbm10
\font\sevenmsb=msbm7
\font\fivemsb=msbm5
\textfont1=\tenmsb
\scriptfont1=\sevenmsb
\scriptscriptfont1=\fivemsb
}

\def\1{{\ds 1}}

\def\C{\hbox{$\bb C$}}

\def\mathbb#1{#1}
\def\mathrm#1{\hbox{#1}}
\def\text#1{{\hbox{#1}}}
\def\mathbf#1{\hbox{\bf #1}}
\def\frac#1#2{{#1\over #2}}
\def\emph#1{{\it #1}}

%\BeniniMZ
\lref\BeniniMZ{
  F.~Benini, Y.~Tachikawa and B.~Wecht,
  ``Sicilian gauge theories and N=1 dualities,''
  arXiv:0909.1327 [hep-th].
  %%CITATION = ARXIV:0909.1327;%%
}

%\IntriligatorAU
\lref\IntriligatorAU{
  K.~A.~Intriligator and N.~Seiberg,
  ``Lectures on supersymmetric gauge theories and electric-magnetic  duality,''
  Nucl.\ Phys.\ Proc.\ Suppl.\  {\bf 45BC}, 1 (1996)
  [arXiv:hep-th/9509066].
  %%CITATION = NUPHZ,45BC,1;%%
}

%\LeighEP
\lref\LeighEP{
  R.~G.~Leigh and M.~J.~Strassler,
  ``Exactly Marginal Operators And Duality In Four-Dimensional N=1
  Supersymmetric Gauge Theory,''
  Nucl.\ Phys.\  B {\bf 447}, 95 (1995)
  [arXiv:hep-th/9503121].
  %%CITATION = NUPHA,B447,95;%%
}

%\KlebanovHH
\lref\KlebanovHH{
  I.~R.~Klebanov and E.~Witten,
  ``Superconformal field theory on threebranes at a Calabi-Yau  singularity,''
  Nucl.\ Phys.\  B {\bf 536}, 199 (1998)
  [arXiv:hep-th/9807080].
  %%CITATION = NUPHA,B536,199;%%
}

%\BenvenutiWI
\lref\BenvenutiWI{
  S.~Benvenuti and A.~Hanany,
  ``Conformal manifolds for the conifold and other toric field theories,''
  JHEP {\bf 0508}, 024 (2005)
  [arXiv:hep-th/0502043].
  %%CITATION = JHEPA,0508,024;%%
}

%\FerraraPZ
\lref\FerraraPZ{
  S.~Ferrara and B.~Zumino,
  ``Transformation Properties Of The Supercurrent,''
  Nucl.\ Phys.\  B {\bf 87}, 207 (1975).
  %%CITATION = NUPHA,B87,207;%%
}

%\DobrevQV
\lref\DobrevQV{
  V.~K.~Dobrev and V.~B.~Petkova,
  ``All Positive Energy Unitary Irreducible Representations Of Extended
  Conformal Supersymmetry,''
  Phys.\ Lett.\  B {\bf 162}, 127 (1985).
  %%CITATION = PHLTA,B162,127;%%
}

%\DixonBG
\lref\DixonBG{
  L.~J.~Dixon,
  ``Some World Sheet Properties of Superstring                   Compactifications, on Orbifolds and Otherwise,''
  Lectures given at the ICTP Summer Workshop in High
Energy Phsyics and Cosmology, Trieste, Italy, Jun 29 -- Aug
7, 1987,
  in {\it Superstrings, unified theories and cosmology}, World Scientific 1988.
  %%CITATION = C87-06-29.1;%%
}

%\AkerblomGX
\lref\AkerblomGX{
  N.~Akerblom, C.~Saemann and M.~Wolf,
  ``Marginal Deformations and 3-Algebra Structures,''
  Nucl.\ Phys.\  B {\bf 826}, 456 (2010)
  [arXiv:0906.1705 [hep-th]].
  %%CITATION = NUPHA,B826,456;%%
}

%\GukovYM
\lref\GukovYM{
  S.~Gukov, E.~Martinec, G.~W.~Moore and A.~Strominger,
  ``The search for a holographic dual to AdS(3) x S**3 x S**3 x S**1,''
  Adv.\ Theor.\ Math.\ Phys.\  {\bf 9}, 435 (2005)
  [arXiv:hep-th/0403090].
  %%CITATION = 00203,9,435;%%
}

%\SeibergPF
\lref\SeibergPF{
  N.~Seiberg,
  ``Observations On The Moduli Space Of Superconformal Field Theories,''
  Nucl.\ Phys.\  B {\bf 303}, 286 (1988).
  %%CITATION = NUPHA,B303,286;%%
}

\lref\OHBK{B.~Kol, unpublished notes.}

%\KolZT
\lref\KolZT{
  B.~Kol,
  ``On conformal deformations,''
  JHEP {\bf 0209}, 046 (2002)
  [arXiv:hep-th/0205141].
  %%CITATION = JHEPA,0209,046;%%
}

%\TachikawaTQ
\lref\TachikawaTQ{
  Y.~Tachikawa,
  ``Five-dimensional supergravity dual of a-maximization,''
  Nucl.\ Phys.\  B {\bf 733}, 188 (2006)
  [arXiv:hep-th/0507057].
  %%CITATION = NUPHA,B733,188;%%
}

%\AsninXX
\lref\AsninXX{
  V.~Asnin,
  ``On metric geometry of conformal moduli spaces of four-dimensional
  %superconformal theories,''
  arXiv:0912.2529 [hep-th].
  %%CITATION = ARXIV:0912.2529;%%
}

%\IntriligatorJJ
\lref\IntriligatorJJ{
  K.~A.~Intriligator and B.~Wecht,
  ``The exact superconformal R-symmetry maximizes a,''
  Nucl.\ Phys.\  B {\bf 667}, 183 (2003)
  [arXiv:hep-th/0304128].
  %%CITATION = NUPHA,B667,183;%%
}

%\GaiottoQI
\lref\GaiottoQI{
  D.~Gaiotto and X.~Yin,
  ``Notes on superconformal Chern-Simons-matter theories,''
  JHEP {\bf 0708}, 056 (2007)
  [arXiv:0704.3740 [hep-th]].
  %%CITATION = JHEPA,0708,056;%%
}

%\ChangSG
\lref\ChangSG{
  C.~M.~Chang and X.~Yin,
  ``Families of Conformal Fixed Points of N=2 Chern-Simons-Matter Theories,''
  arXiv:1002.0568 [hep-th].
  %%CITATION = ARXIV:1002.0568;%%
}

%\BianchiJA
\lref\BianchiJA{
  M.~S.~Bianchi, S.~Penati and M.~Siani,
  ``Infrared stability of ABJ-like theories,''
  JHEP {\bf 1001}, 080 (2010)
  [arXiv:0910.5200 [hep-th]].
  %%CITATION = JHEPA,1001,080;%%
}
%\BianchiRF
\lref\BianchiRF{
  M.~S.~Bianchi, S.~Penati and M.~Siani,
  ``Infrared Stability of N=2 Chern-Simons Matter Theories,''
  arXiv:0912.4282 [hep-th].
  %%CITATION = ARXIV:0912.4282;%%
}

%\StromingerPD
\lref\StromingerPD{
  A.~Strominger,
  ``Special Geometry,''
  Commun.\ Math.\ Phys.\  {\bf 133}, 163 (1990).
  %%CITATION = CMPHA,133,163;%%
}

%\PapadodimasEU
\lref\PapadodimasEU{
  K.~Papadodimas,
  ``Topological Anti-Topological Fusion in Four-Dimensional Superconformal
  %Field Theories,''
  arXiv:0910.4963 [hep-th].
  %%CITATION = ARXIV:0910.4963;%%
}

%\MinahanFG
\lref\MinahanFG{
  J.~A.~Minahan and D.~Nemeschansky,
  ``An N = 2 superconformal fixed point with $E_6$ global symmetry,''
  Nucl.\ Phys.\  B {\bf 482}, 142 (1996)
  [arXiv:hep-th/9608047].
  %%CITATION = NUPHA,B482,142;%%
}

%\AharonyHX
\lref\AharonyHX{
  O.~Aharony, B.~Kol and S.~Yankielowicz,
  ``On exactly marginal deformations of $\cN = 4$ SYM and type IIB  supergravity on
  AdS$_5$ $\times$ S$^5$,''
  JHEP {\bf 0206}, 039 (2002)
  [arXiv:hep-th/0205090].
  %%CITATION = JHEPA,0206,039;%%
}

%\KutasovXB
\lref\KutasovXB{
  D.~Kutasov,
  ``Geometry on the space of conformal field theories and contact terms,''
  Phys.\ Lett.\  B {\bf 220}, 153 (1989).
  %%CITATION = PHLTA,B220,153;%%
}

%\KutasovUX
\lref\KutasovUX{
  D.~Kutasov,
  ``New results on the `a-theorem' in four dimensional supersymmetric field
  theory,''
  arXiv:hep-th/0312098.
  %%CITATION = HEP-TH/0312098;%%
}

%\ArgyresXU
\lref\ArgyresXU{
  P.~C.~Argyres, K.~A.~Intriligator, R.~G.~Leigh and M.~J.~Strassler,
  ``On inherited duality in N = 1 d = 4 supersymmetric gauge theories,''
  JHEP {\bf 0004}, 029 (2000)
  [arXiv:hep-th/9910250].
  %%CITATION = JHEPA,0004,029;%%
}

%\BarnesJJ
\lref\BarnesJJ{
  E.~Barnes, K.~A.~Intriligator, B.~Wecht and J.~Wright,
  ``Evidence for the strongest version of the 4d a-theorem, via  a-maximization
  along RG flows,''
  Nucl.\ Phys.\  B {\bf 702}, 131 (2004)
  [arXiv:hep-th/0408156].
  %%CITATION = NUPHA,B702,131;%%
}

%\ArgyresCN
\lref\ArgyresCN{
  P.~C.~Argyres and N.~Seiberg,
  ``S-duality in N=2 supersymmetric gauge theories,''
  JHEP {\bf 0712}, 088 (2007)
  [arXiv:0711.0054 [hep-th]].
  %%CITATION = JHEPA,0712,088;%%
}

%\NelsonMQ
\lref\NelsonMQ{
  A.~E.~Nelson and M.~J.~Strassler,
  ``Exact results for supersymmetric renormalization and the supersymmetric
  flavor problem,''
  JHEP {\bf 0207}, 021 (2002)
  [arXiv:hep-ph/0104051].
  %%CITATION = JHEPA,0207,021;%%
}

%\SeibergVC
\lref\SeibergVC{
  N.~Seiberg,
  ``Naturalness Versus Supersymmetric Non-renormalization Theorems,''
  Phys.\ Lett.\  B {\bf 318}, 469 (1993)
  [arXiv:hep-ph/9309335].
  %%CITATION = PHLTA,B318,469;%%
}

%\FreedmanGP
\lref\FreedmanGP{
  D.~Z.~Freedman, S.~S.~Gubser, K.~Pilch and N.~P.~Warner,
  ``Renormalization group flows from holography supersymmetry and a
  c-theorem,''
  Adv.\ Theor.\ Math.\ Phys.\  {\bf 3}, 363 (1999)
  [arXiv:hep-th/9904017].
  %%CITATION = 00203,3,363;%%
}

%\DixonFJ
\lref\DixonFJ{
  L.~J.~Dixon, V.~Kaplunovsky and J.~Louis,
  ``On Effective Field Theories Describing (2,2) Vacua of the Heterotic
  String,''
  Nucl.\ Phys.\  B {\bf 329}, 27 (1990).
  %%CITATION = NUPHA,B329,27;%%
}

%\MeadeWD
\lref\MeadeWD{
  P.~Meade, N.~Seiberg and D.~Shih,
  ``General Gauge Mediation,''
  Prog.\ Theor.\ Phys.\ Suppl.\  {\bf 177}, 143 (2009)
  [arXiv:0801.3278 [hep-ph]].
  %%CITATION = PTPSA,177,143;%%
}

%\LucchesiIR
\lref\LucchesiIR{
  C.~Lucchesi and G.~Zoupanos,
  ``All-order Finiteness in $\cN=1$ SYM Theories: Criteria and Applications,''
  Fortsch.\ Phys.\  {\bf 45}, 129 (1997)
  [arXiv:hep-ph/9604216].
  %%CITATION = FPYKA,45,129;%%
}

%\SeibergPQ
\lref\SeibergPQ{
  N.~Seiberg,
  ``Electric -- magnetic duality in supersymmetric nonAbelian gauge theories,''
  Nucl.\ Phys.\  B {\bf 435}, 129 (1995)
  [arXiv:hep-th/9411149].
  %%CITATION = NUPHA,B435,129;%%
}

%\ZamolodchikovGT
\lref\ZamolodchikovGT{
  A.~B.~Zamolodchikov,
  ``Irreversibility of the Flux of the Renormalization Group in a 2D Field
  Theory,''
  JETP Lett.\  {\bf 43}, 730 (1986)
  [Pisma Zh.\ Eksp.\ Teor.\ Fiz.\  {\bf 43}, 565 (1986)].
  %%CITATION = ZFPRA,43,565;%%
}

%\CecottiQN
\lref\CecottiQN{
  S.~Cecotti, S.~Ferrara and L.~Girardello,
  ``Geometry of Type II Superstrings and the Moduli of Superconformal Field
  Theories,''
  Int.\ J.\ Mod.\ Phys.\  A {\bf 4}, 2475 (1989).
  %%CITATION = IMPAE,A4,2475;%%
}

%\CecottiZX
\lref\CecottiZX{
  S.~Cecotti, S.~Ferrara and L.~Girardello,
  ``A topological formula for the K\"ahler potential of 4d $\cN=1$, $\cN=2$ strings and its implications for the moduli problem,''
  Phys.\ Lett.\  B {\bf 213}, 443 (1988).
  %%CITATION = PHLTA,B213,443;%%
}

%\FlatoTE
\lref\FlatoTE{
  M.~Flato and C.~Fronsdal,
  ``Representations Of Conformal Supersymmetry,''
  Lett.\ Math.\ Phys.\  {\bf 8}, 159 (1984).
  %%CITATION = LMPHD,8,159;%%
}

%\MinwallaKA
\lref\MinwallaKA{
  S.~Minwalla,
  ``Restrictions imposed by superconformal invariance on quantum field
  theories,''
  Adv.\ Theor.\ Math.\ Phys.\  {\bf 2}, 781 (1998)
  [arXiv:hep-th/9712074].
  %%CITATION = 00203,2,781;%%
}

%\OsbornQU
\lref\OsbornQU{
  H.~Osborn,
  ``$\cN = 1$ superconformal symmetry in four-dimensional quantum field theory,''
  Annals Phys.\  {\bf 272}, 243 (1999)
  [arXiv:hep-th/9808041].
  %%CITATION = APNYA,272,243;%%
}

%\NovikovUC
\lref\NovikovUC{
  V.~A.~Novikov, M.~A.~Shifman, A.~I.~Vainshtein and V.~I.~Zakharov,
  ``Exact Gell-Mann-Low Function Of Supersymmetric Yang-Mills Theories From
 Instanton Calculus,''
  Nucl.\ Phys.\  B {\bf 229}, 381 (1983).
  %%CITATION = NUPHA,B229,381;%%
}

%\BarnesBW
\lref\BarnesBW{
  E.~Barnes, E.~Gorbatov, K.~A.~Intriligator and J.~Wright,
  ``Current correlators and AdS/CFT geometry,''
  Nucl.\ Phys.\  B {\bf 732}, 89 (2006)
  [arXiv:hep-th/0507146].
  %%CITATION = NUPHA,B732,89;%%
}

%\StrasslerIZ
\lref\StrasslerIZ{
  M.~J.~Strassler,
  ``On renormalization group flows and exactly marginal operators in three
  dimensions,''
  arXiv:hep-th/9810223.
  %%CITATION = HEP-TH/9810223;%%
}

\def\cP{{\cal P}}
\def\cN{{\cal N}}
\def\cM{{\cal M}}
\def\cO{{\cal O}}

\def\cL{{\cal L}}
\def\Tr{\mathop{\mathrm{Tr}}}

\def\Im{\mathop{\mathrm{Im}}}

\def\br{\mathbf{r}}
\def\cW{{\cal W}}
\def\vev#1{\langle#1\rangle}
\def\SCFT{\text{SCFT}}
\def\cO{{\cal O}}
\def\cN{{\cal N}}
\def\cW{{\cal W}}
\def\half{{1\over 2}}

\def\zlambda{\lambda'\,{}}

%\draftmode

\Title{}{Exactly Marginal Deformations and  Global Symmetries}

\bigskip

\centerline{Daniel Green, Zohar Komargodski, Nathan Seiberg,}
\centerline{Yuji Tachikawa, and Brian Wecht\footnote{$^\dagger$}{drgreen, zoharko, seiberg, yujitach, bwecht@ias.edu}}

\bigskip

\centerline{School of Natural Sciences,
Institute for Advanced Study,}
\centerline{Einstein Drive, Princeton, NJ 08540, USA}

\bigskip
\bigskip
\bigskip

\noindent
We study the problem of finding exactly marginal deformations of $\cN=1$ superconformal field theories in four dimensions.  We find that the only way a marginal chiral operator can become not exactly marginal is for it to combine with a conserved current multiplet. Additionally, we find that the space of exactly marginal deformations, also called the ``conformal manifold," is the quotient of the space of marginal couplings by the complexified continuous global symmetry group. This fact explains why exactly marginal deformations are ubiquitous in $\cN=1$ theories.  Our method turns the problem of enumerating exactly marginal operators into a problem in group theory, and substantially extends and simplifies the previous analysis by Leigh and Strassler.  We also briefly discuss how to apply our analysis to $\cN=2$ theories in three dimensions.

\Date{May 2010}

\eject

\newsec{Introduction}

Conformal field theories (CFTs) play a broad role in understanding quantum field theory and its applications, as they often allow exact results which can be difficult to obtain in massive theories. In a given CFT, operators are classified as irrelevant, marginal, or relevant according to their scaling dimensions. Deformations by these operators then control the renormalization group (RG) flow close to the conformal fixed point. It is especially interesting to perturb the theory by a marginal operator, since this deformation preserves conformality at zeroth order. However, since the dimension of the deforming operator is often itself corrected, we can further subdivide the marginal operators into those which are marginally relevant, marginally irrelevant, or exactly marginal. In many cases, it is difficult to tell to which of these three classes a marginal operator belongs. It is the goal of the present work to explore this question in supersymmetric theories.

When an operator is exactly marginal, one can perturb the CFT without breaking conformal invariance. This perturbation then gives rise to a family of CFTs near the original fixed point.  If there are multiple such operators, one can then locally think of the extended family of CFTs as a manifold in the space of couplings. This manifold is conventionally called the ``conformal manifold,'' which we denote by $\cM_c$.  This space has a natural metric derived by the two point function of the marginal operators \ZamolodchikovGT.

The conformal manifold of two-dimensional CFTs is especially interesting, as it becomes the space of vacua when the CFTs are used as the internal part of the worldsheet CFT of a string theory. As a result, conformal manifolds for two-dimensional theories have been extensively studied, especially in cases where the theories have $\cN=(2,2)$ supersymmetry.\foot{An analysis mostly from the two-dimensional point of view was given in~\DixonBG, while an analysis from the spacetime point of view was initiated in~\SeibergPF.  Additionally, the geometry of the conformal manifold was further studied in \KutasovXB.  These studies  led to the understanding of the special geometry of conformal manifolds of $c=9$ $\cN=(2,2)$ theories \refs{\CecottiZX\CecottiQN\DixonFJ{--}\StromingerPD}.}
In these theories, marginal operators correspond to massless spacetime fields, and exactly marginal operators lead to the moduli space of vacua.

The conformal manifold of four-dimensional $\cN=1$ superconformal field theories (SCFTs) was first explored by Leigh and Strassler in \LeighEP.  In this work, the authors describe how the beta functions of gauge and superpotential couplings can be linearly dependent, and how this dependence implies the existence of exactly marginal operators. With the advent of the AdS/CFT correspondence, conformal manifolds of boundary CFTs could be mapped to the vacua of AdS theories. Simple facts about the vacua of AdS supergravity then provide insight into the structure of the conformal manifold \refs{\AharonyHX\KolZT\TachikawaTQ{--}\BarnesBW}. For example, we expect the conformal manifold to be not only K\"ahler,\foot{Indeed, this has been proven in~\AsninXX.} but also to arise from a Higgs mechanism where the bulk vector fields (which correspond to the global symmetries on the boundary) ``eat'' some of the bulk scalar fields (which correspond to the deformations of the boundary theory).

Our aim in this note is to establish the above properties of $\cM_c$ directly by using only field-theoretic techniques. As a consequence, our results will be valid for any field theories with conformal manifolds, even when there is no dual gravity description. In particular, we show that when a given superconformal field theory with global continuous (non-R) symmetry group $G$ is deformed by a marginal superpotential $\cW = \lambda^i \cO_i$ (where $\lambda^i$ are couplings and $\cO_i$ are marginal operators),  the conformal manifold is given (in some small but finite neighborhood of the theory with $\lambda^i=0$) by dividing the space of couplings by the complexified symmetry group, $\cM_c = \{ \lambda^i \} / G^{\C}$.

The importance of the global continuous symmetry group $G$ is highlighted by the following fact.  Consider the conformal field theory at a point $\cP$ on $\cM_c$ (we will refer to this theory as $\cP$), where a particular marginal operator is not exactly marginal.  We show that such an operator is marginally irrelevant at $\cP$ and that it is irrelevant at generic points on $\cM_c$.  More explicitly, as we move away from $\cP$ and the dimension of this operator is lifted, the operator can no longer be in a short multiplet. Instead, it is lifted by pairing with another operator in a short multiplet $J_a$.  This operator $J_a$ is a conserved current associated with a generator of $G$ which is unbroken at $\cP$ but is broken elsewhere.

As part of our proof, we will demonstrate that any marginal operator invariant under the global symmetry group must be exactly marginal. Our analysis also explains why we often find exactly marginal deformations in $\cN=1$ SCFTs.
Some of our statements have previously appeared in various forms in the literature~\refs{\KolZT,\OHBK\ArgyresXU-\NelsonMQ}.  In particular, \refs{\KolZT,\OHBK} have conjectured the connection between the exactly marginal operators and the $D$-term equations.  Our goal here is to provide a rigorous derivation from field theory.

One important aspect of our method is that it does not use the NSVZ beta function~\NovikovUC\ and almost solely relies on the $\cN=1$ superconformal algebra. Therefore, it applies to any superconformal field theory with or without a Lagrangian description.  It also applies to three-dimensional $\cN=2$ SCFTs, which share most of the multiplet structures of four-dimensional $\cN=1$ SCFTs.

As in~\SeibergVC, we can promote all the coupling constants to background superfields.  The couplings $\lambda^i$ of chiral operators $\cO_i$ are promoted to background chiral superfields and the couplings $Z^a$ of conserved currents $J_a$ are promoted to background vector superfields.  Then we find that at least to leading order around the conformal manifold $\cM_c$, the renormalization group flow is a gradient flow based on an action on the space of coupling constants.  The symplectic quotient which determines $\cM_c$ is related to the $D$-term equations of these background superfields.  A similar picture in terms of an action in the space of coupling constants is natural in the worldsheet description of string theory and is crucial in the AdS/CFT correspondence.  Here we see that it arises in every $\cN=1$ superconformal theory.

The outline of the paper is as follows.  In Sec.~2, we begin by  presenting a general analysis of the structure of $\cM_c$, starting with a brief introduction and setup in Sec.~2.1. In particular, in Sec.~2.2, we describe $\cM_c$ as a symplectic quotient; we then relate this description to the renormalization group flow in Sec.~2.3. When incorporating the free limit of gauge fields, our argument has a subtlety which we deal with in Sec.~2.4.  In Sec.~2.5, we very briefly describe how to apply our analysis to $\cN=2$ supersymmetry in three dimensions.  In Sec.~3, we give a point-by-point comparison of our method with that of Leigh and Strassler.  In Sec.~4,
we discuss five explicit examples which illustrate the salient features of our methods. Finally, in Sec.~5, we conclude with a short discussion of future directions.

\newsec{Geometry of the Conformal Manifold}

\subsec{Setup}

Suppose we are given a superconformal theory, which we will call $\cP$. $\cP$ could be somewhere along a line of fixed points including a free field theory (Fig.~1) or an IR fixed point of an asymptotically free theory (Fig.~2).  The precise nature of the construction of $\cP$ can help enumerate operators of this theory, but for our purposes here it is unimportant how $\cP$ is obtained.

In this work, we will be interested in supersymmetric deformations of the theory.  Depending on their dimensions these deformations can be relevant, marginal, or irrelevant.
In this work we will only concern ourselves with marginal deformations.

Using the superconformal algebra, we show in Appendix A that there are only two kinds of supersymmetric deformations.  These deformations can be described as chiral operators which are integrated over half of superspace or generic operators which are integrated over all of superspace.  We will refer to these two types as superpotential deformations and K\"ahler deformations, respectively, in analogy with the terminology used when $\cP$ is a free theory.  For the superpotential deformations to be marginal, we need chiral primary operators $\cO_i$ of dimension three.  For the K\"ahler deformations to be marginal, we need real primary operators $J_a$ of dimension two.  Such $J_a$ are conserved currents, satisfying the conservation equation
\eqn\conscu{\bar D^2 J_a=0 ~.}
Therefore  $\int d^4 \theta J_a = 0$, and thus $J_a$ does not deform the Lagrangian.  Hence there are no marginal K\"ahler deformations.  We will, however, soon see that the set of conserved currents $J_a$ still plays an important role in determining the structure of the conformal manifold.

\topinsert
\tabskip=0em plus1fil
\halign to\hsize{  #  \hfil
\tabskip=0em plus2fil& # \hfil
\tabskip=0em plus1fil \cr
\hfill\epsfxsize5.5cm \epsffile{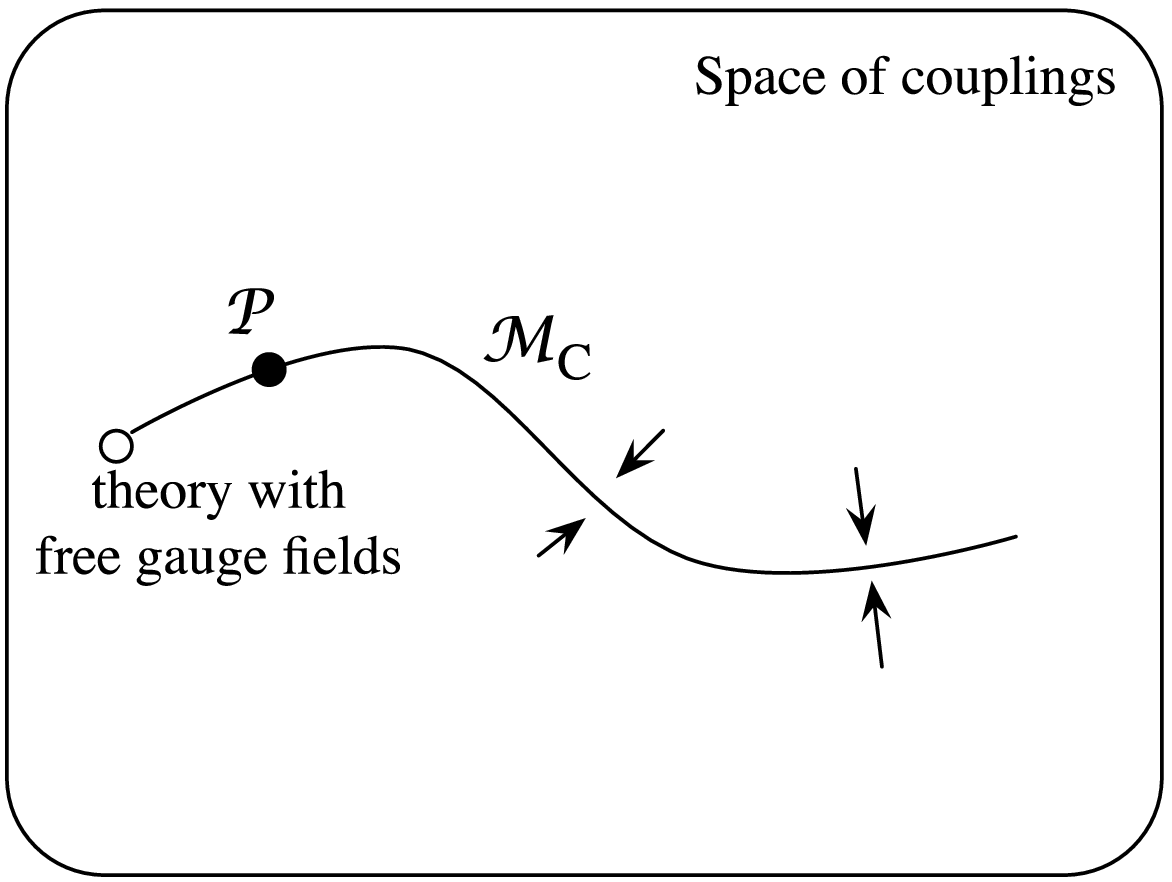}\hfill &
\hfill\epsfxsize5.5cm\epsffile{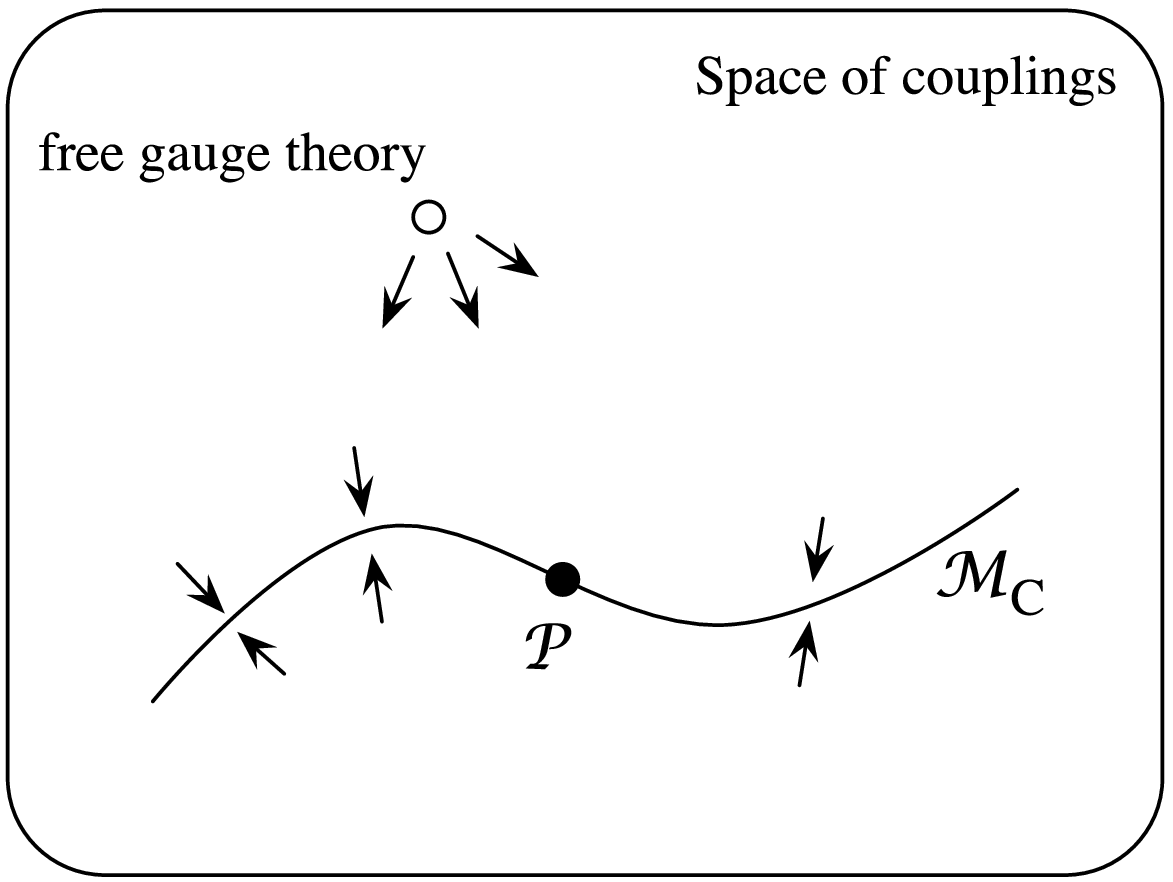} \hfill\cr
Fig. 1:  $\cP$ could be along a line of&
Fig. 2:  $\cP$ could be the IR fixed point \cr
fixed points  including a free gauge &
of  a gauge theory,  as in $\cN=1$ SQCD.\cr
theory, as in $\cN=4$ SYM.&\cr}
\bigskip
\endinsert

It will be useful for our purposes here to note that at $\cP$, two-point functions determine natural Zamolodchikov metrics $g_{i\bar\jmath}$ and $\gamma_{ab}$ which can be used to raise and lower indices. The relevant two-point functions and associated metrics are given by
\eqn\tmpeqnlabel{
\langle \cO_i(x) \cO^\dagger_{\bar\jmath}(0) \rangle = {g_{i\bar \jmath} \over |x|^6}\qquad {\rm and} \qquad
\langle J_a(x) J_b(0) \rangle = {\gamma_{ab} \over |x|^4}~.
}

\subsec{Identifying the conformal manifold}

We consider  deforming $\cP$ by
\eqn\super{
\cW = \lambda^i \cO_i~}
where $\cO_i$ are chiral operators of dimension three and $\lambda^i$ are small but finite coefficients.  For now  we assume that there are no free gauge fields, because turning on a small gauge coupling for them does not correspond to a small superpotential deformation. We will come back to this case in Sec.~2.4.  We would like to determine for which  $\lambda^i$ the theory remains conformal. Collectively, such exactly marginal directions comprise the conformal manifold ${\cal M}_c$ in a neighborhood of $\cP$.

It is important that there are no singular terms in the OPE of two chiral operators.  Therefore, we can use a renormalization scheme in which the superpotential \super\ is not renormalized.  In this scheme, only the K\"ahler potential can be modified.  Since the lowest dimension operators which can appear in the K\"ahler potential are the currents $J_a$,
we can have
\eqn\change{
\cL \to \cL + \int d^4\theta  Z^a(\lambda,\bar\lambda ;\mu) J_a~,
}
where $\mu$ is the distance cutoff in conformal perturbation theory. Operators of higher dimension are irrelevant and thus we do not consider them.  We should stress that $\int d^4 \theta J_a$ vanishes at $\cP$, but this is no longer the case once the perturbation \super\ is included.  We will describe this situation in more detail below.

Right away, we note that if there are no such operators $J_a$ in $\cP$ (that is, if the original theory has no global continuous non-R symmetries), there can be no renormalization. In this case, all marginal operators are in fact exactly marginal. This simple result is illustrative of the power of the argument we employ in this work.

When there are global symmetry currents, the condition for the deformed theory to be superconformal is that~\change\ is independent of the cutoff
\eqn\dzdef{
D^a(\lambda,\bar\lambda)\equiv \mu\frac{\partial}{\partial \mu} Z^a(\lambda,\bar\lambda;\mu) =0~.
}
Furthermore, we need to identify deformations related by the global symmetries $G$ because the resulting conformal field theories are identical.  Thus the conformal manifold close to $\cP$ is a quotient
\eqn\tmpeqnlabelXXX{\cM_c= \{\lambda^i | D^a=0\} / G~.}

In order to gain further insight into the objects $D^a$ in \dzdef, we now examine them in conformal perturbation theory. To lowest non-trivial order $Z^a(\lambda,\bar\lambda; \mu) $ is determined by the operator product expansion
\eqn\opecac{\cO_i(x) \cO_{\bar\jmath}^\dagger(0) = {g_{i\bar\jmath} \over |x|^6} + {T^a_{i\bar \jmath} \over |x|^4}J_a(0)+\cdots}
Here $g_{i\bar\jmath} $ is the metric in \tmpeqnlabel\ and the coefficients $T^a_{i\bar \jmath}$ are a representation of the global symmetry group (when it is Abelian, $T^a_{i\bar\jmath}=q_b^i \gamma^{ab} g_{i \bar j}$, where $q_b^i$ is the charge of the operator $\cO_i$). Equivalently, we can say that
\eqn\threept{
\vev{\cO_i(x) \cO_{\bar\jmath}^\dagger(y) J_a(z)} = \frac{\gamma_{ab} T^b_{i\bar\jmath}}{|x-y|^{4} |x-z|^2 |y-z|^2} ~.}
This leads to a logarithmic singularity in
\eqn\oobarint{
\left(\int d^4x\ d^2 \theta \lambda^i \cO_i(x)\right) \left(\int d^2 \bar \theta  \,  \bar\lambda^{\bar\jmath} \cO^\dagger_{\bar\jmath} (0) \right) \sim \int d^4 \theta Z^a(\lambda,\bar\lambda; \mu ) J_a(0) ~.}
(Note that the identity operator in \opecac\ does not contribute to \oobarint\ and the integral of $J_a$ is nonzero only at the next order in conformal perturbation theory.)
Comparing \threept\ with \oobarint\ and \dzdef, we identify
\eqn\D{
D^a=   2 \pi^2 \lambda^i \, T_{i\bar\jmath}^a \, \bar\lambda^{\bar\jmath}+\cdots
}
We recognize the leading order expression in \D\ as the moment map of the global symmetry $G$ acting on the space of $\lambda^i$.  Higher-order corrections do not affect our conclusions in the vicinity of $\cP$ -- they only change the value of $D^a$.

Thus far, we have employed a holomorphic renormalization scheme where the only change in the Lagrangian is the deformation \change\ of the K\"ahler potential. We can, however, usefully rephrase this change as a renormalization group flow of the superpotential couplings $\{\lambda^i\}$ by using the non-conservation of the current.

The deformation~\super\ in general breaks the global symmetry.  In the deformed theory, we have the equation
\eqn\non{
\bar D^2 J_a = X_a^i \cO_i~ \qquad
\text{where}\qquad
X_a^i= \lambda^j\, T^{i}_{j a}   + \cdots ~.
}
Here, $X_a^i$ is the vector field on the space of $\{\lambda^i\}$ representing the action of $J_a$, and its leading order form follows from \threept.
We can therefore rewrite the deformation \change\  as the change of $\lambda^i$ instead:
 \eqn\tmpeqnlabelXXXXX{
\lambda^i \to \lambda^i - \frac{1}{2} X_a^i Z^a~.
}
Note that this change is not holomorphic. The beta function of $\lambda^i$ is given by
 \eqn\tmpeqnlabelXXXXXX{
\mu\frac{\partial}{\partial \mu} \lambda^i = - \frac{1}{2} X_a^i D^a~.
}
To leading order, we find
\eqn\tmpeqnlabelXXXXXXX{
\mu \frac{\partial}{\partial\mu} \lambda^i
=   -  \pi^2 (\lambda^j T_{j a}^i) ( \lambda^l T_{l\bar m}^b \bar\lambda^{\bar m}) + \dots
= - (8 \pi^2)^{-1} g^{i\bar\jmath} \frac{\partial}{\partial \bar\lambda^{\bar\jmath}}(\gamma_{ab} D^a D^b ) + \dots
}
Note that this flow is a gradient flow generated by $\gamma_{ab}D^a D^b$.
Since $\gamma_{ab}$ is positive definite, $\gamma_{ab}D^a D^b \ge 0$, with the inequality saturated when $D^a=0$.

The higher order corrections to \tmpeqnlabelXXXXXXX\ do not affect our qualitative conclusions.  As in our discussion around \change\ -- \tmpeqnlabelXXX, the only way conformal invariance can break is if $D^a$ is nonzero.  Starting with a solution of the leading order equation $D^a=0$, we can correct it order by order in a power series in $\lambda$ to find a solution of the full equations.

Our perturbative computation allows us to explore the vicinity of the locus $\{\lambda^i | D^a=0\}$.  It tells us that all the transverse directions to this locus correspond to (marginally) irrelevant operators; i.e.\ $\{\lambda^i | D^a=0\}$ is attractive.\foot{Recall that typically there are some relevant superpotential deformations which we do not discuss here.} Said another way, we have shown that marginal deformations are either exactly marginal or marginally irrelevant, but never marginally relevant. Additionally, we have discovered that to leading order, the renormalization group flow is a gradient flow.

Finally, the symmetry group $G$ which acts on $\{\lambda^i | D^a=0\}$ does not affect the conformal field theory.  Therefore, the space of superconformal field theories is the quotient \tmpeqnlabelXXX\ which can also be written as
\eqn\hq{\cM_c = \{ \lambda^i \} / G^{\C}~.}

\subsec{Symplectic quotient and renormalization group}

We concluded above that the conformal manifold is a symplectic quotient.
In this subsection we give an interpretation of this quotient.

The flow induced by a small deformation $\cW=\lambda^i \cO_i$ leads to an infrared SCFT close to $\cP$; we denote the resulting theory by $\SCFT[\lambda^i]$.  We now wish to understand the circumstances under which $\SCFT[\lambda^i]$ and $\SCFT[\zlambda^i]$  are identical for  $\lambda^i\ne\zlambda^i$.
In other words, we wish to introduce an equivalence relation
$\lambda^i \sim \zlambda^i$ by means of the RG flow.
The conformal manifold is then $\cM_c=\{\lambda^i\} / \sim$.
We will  soon see that this identification by the RG flow
is the identification by the complexified global symmetry group.

For simplicity, we assume  that $G=U(1)$ (the extension to any symmetry group $G$ is trivial) and that the space of marginal operators is spanned by $\cO_i$ with charges $q^i$.  Consider the deformation by small but finite $\cW=\lambda^i\cO_i$, which leads to $\SCFT[\lambda^i]$.
Alternately, we could deform the theory by $\cW=\lambda^i e^{\epsilon q^i} \cO_i$ for some small complex $\epsilon$; the resulting theory is  $\SCFT[\lambda^i e^{\epsilon q^i}]$. We assume that both $\SCFT[\lambda^i]$ and $\SCFT[\lambda^i e^{\epsilon q^i}]$ are not equivalent to $\cP$; if so, we can revert to the perturbative calculation of the previous subsection.
The difference between the two theories to leading order is the operator
\eqn\thydif{
\delta \cL= \epsilon\int d^2\theta\sum_i q^i\lambda^i \cO_i+ c.c.+ \cO(\epsilon^2).
}
The $U(1)$ current is broken and satisfies
$\bar{D}^2 J=\sum_i q^i\lambda^i \cO_i$.
As a result, $\Delta[q^i\lambda^i\cO_i]>3$ because this operator is no longer a primary at $\SCFT[\lambda^i]$.
Therefore, the two theories differ by an irrelevant operator, and thus flow to the same infrared theory.
The result is that to describe the conformal manifold, we need to impose identification of the $\lambda^i$ under the action of the complexified symmetry group $G^{\C} $.

We can easily prove the converse; $\SCFT[\lambda^i]$ and $\SCFT[\zlambda^i]$ are only equivalent when $\lambda^i$ and $\zlambda^i$ differ by a complexified symmetry transformation.
If $\SCFT[\lambda^i]$ = $\SCFT[\zlambda^i]$ for very close $\lambda^i$ and $\zlambda^i$, the theories must differ by a real irrelevant operator whose dimension is arbitrarily close to $2$.  For our $U(1)$ case, the only such operator is $J_a$, and so the $\lambda$s must be related by a complexified symmetry transformation.

\subsec{With free gauge fields}

The analysis so far is not directly applicable when the reference point $\cP$ contains a free non-Abelian gauge multiplet $W_\alpha$.  This is because the perturbation in the gauge coupling $g$ around the free theory is not given by a small addition of a gauge-invariant operator $\Tr W_\alpha W^\alpha$ to the superpotential. In this section, we describe how to modify our argument to include this case.  Note that the analysis in the previous section already covered the case in Fig.~2, when the gauge coupling at $\cP$ is nonzero but small.  Here we work directly at the point on the space of couplings where there are free gauge fields.

Suppose $\cP$ consists of a free gauge field $W_\alpha$ of the gauge group $G$, and also a ``matter theory'' with flavor symmetry $F$.  The matter theory can be either a theory of free chiral multiplets, or a strongly coupled SCFT.  We gauge a $G$ subgroup of $F$.  Call $H$ the maximal subgroup of  $F$ which commutes with $G$, so $F\supset G\times H$. Note that some part of $H$ can be anomalously broken by the coupling of the matter theory to $G$; this fact will be important for our analysis.

Let $\cO_i$ be the set of $G$-invariant chiral marginal operators of the matter theory, and let us turn on a small gauge coupling and also a small superpotential
\eqn\tmpeqnlabelXXXXXXXX{
\cW=\lambda^i \cO_i  + \frac{\tau}{8 \pi i } \Tr W_{\alpha} W^{\alpha}.
}
where $\tau = {\Theta}/{2 \pi} + i {4 \pi}/{g^2}$.  We take the holomorphic renormalization scheme so that $\lambda^i$ is unchanged, while $\tau$ runs only at one loop. We can further assume that the two-point function of the current of $G$ of the matter theory is such that there is no one-loop running of $\tau$; otherwise, we are in the situation of Fig. 2.

As in \change, quantum effects  change the K\"ahler potential.  Gauge invariance dictates that $J_a$ are currents of $H$. We can now follow the previous argument  almost verbatim, by replacing the set of couplings $\{\lambda^i\}$ by $\{\tau,\lambda^i\}$.  Again, it is important to remember that the anomalous part of $H$ acts not only on $\lambda^i$ but also on $\tau$. We conclude that $\cM_c=\{\tau,\lambda^i\}/H^{\C}$.

We can check this general analysis by a perturbative calculation in $\lambda^i$ and $g$.  This calculation is standard perturbation theory when the matter theory is a theory of free chiral multiplets; if not, it is a mixture of weak gauging of a flavor symmetry of a strongly coupled sector as in \refs{\ArgyresCN,\MeadeWD}  and conformal perturbation theory.  We find
\eqn\DwithGauge{
D^a= 2 \pi^2 \lambda^i T^a_{i\bar\jmath} \bar\lambda^{\bar\jmath} - k^a g^2+ \dots = 2\pi^2 \lambda^i T^a_{i\bar\jmath} \bar\lambda^{\bar\jmath} - k^a 4\pi (\Im\tau)^{-1}+\dots
}
Here, $k^a$ is determined by the three-point function \eqn\tmpeqnlabelXXXXXXXXX{
\vev{J_a J_AJ_B} = \frac{k_a \gamma_{AB}}{|x-y|^2|x-z|^2 |y-z|^2} ~,
}
where $J_a$ is the current of $H$ and $J_{A,B}$ are the currents of $G$.  The coefficient $k_a$ also enters in the anomalous conservation of the current $J_a$ via
\eqn\anomalousconservation{
\bar{D}^2 J_a = \lambda^i T^{j}_{ia}  \cO_j  + k_a \Tr W_{\alpha}W^{\alpha}~.
}

Note that at weak coupling $\vev{\Tr WW(x) \Tr WW(0)}\sim (\Im\tau)^{-2}/|x|^6$.  Therefore it is natural to introduce the metric $g_{\tau\bar\tau}\sim(\Im\tau)^{-2}$ on the space of $\tau$. Then it is easy to see that $k^a (\Im\tau)^{-1}$ is the moment map for the anomalous shift $\tau\to \tau+k^a$, and $D^a$ above is the total moment map acting on the space of $\{\tau,\lambda^i\}$.  Equivalently, this statement means that $e^{2\pi i \tau}$ has charge $k_a$ under the anomalous symmetry \IntriligatorAU.

Using \anomalousconservation, we can again rewrite $Z^aJ_a$ as a non-holomorphic change of the superpotential, and obtain
\eqn\betafunction{
\eqalign{\mu \frac{\partial}{\partial\mu} \lambda^i
&=  - \frac{1}{2} (\lambda^j T_{j\ a}^i) D^a
= -(8 \pi^2)^{-1} g^{i\bar\jmath} \frac{\partial}{\partial \bar\lambda^{\bar\jmath}}(\gamma_{ab} D^a D^b )+\dots \cr
\mu \frac{\partial}{\partial\mu} \tau
&= -i 4 \pi  k_a D^a
= - g^{\tau\bar\tau} \frac{\partial}{\partial \bar\tau}(\gamma_{ab} D^a D^b )+\dots }}
Note that this is again a gradient flow generated by $\gamma_{ab}D^a D^b$.
Therefore, up to this order, we identify the conformal manifold as
$\cM_c=\{\tau,\lambda^i\} / H^{\C}.$

It is instructive to compare this general analysis to the standard case  of gauge theories with vanishing one-loop beta functions \LucchesiIR. Consider a gauge multiplet of group $G$  coupled to matter fields $Q_a$ transforming in $\br_a$, such that $3\,t(\mathbf{adj})=\sum_a t(\br_a)$, where $t(\br)$ is the quadratic Casimir of $\br$.  For simplicity let us further take $\lambda^i=0$.
It is well-known that  $(\mu\partial/\partial\mu) (1/g^2) \sim g^2 \sum_a t(\br_a)^2/|\br_a|$. This is in accord with \betafunction. Indeed,  we have one anomalous $U(1)$ current $J_a=Q^\dagger_a Q_a$ for each irreducible multiplet $\br_a$. In this normalization, we have $\gamma_{ab}=|\br_a| \delta_{ab}$ and $k_a=t(\br_a)$.

\subsec{$\cN=2$ supersymmetry in three dimensions}

Our analysis is readily generalizable to $\cN=2$ superconformal theories in three dimensions, because the structure of the multiplets is quite similar to that of the four-dimensional $\cN=1$ case.
The existence of the conformal manifold in $\cN=2$ Chern-Simons-matter theories was found in \refs{\GaiottoQI\AkerblomGX\BianchiJA-\BianchiRF} by an explicit calculation and an all-orders argument for  weakly-coupled  theories was given in \ChangSG. Here we extend the analysis to the strongly-coupled case.\foot{See also \StrasslerIZ\ and Sec.~5 of \KolZT\ for early works in this direction.}

In three dimensions, the marginal superpotential deformation is given by
dimension-two chiral primary operators $\cO_i$.  The real primary operators saturate the unitarity bound when they are conserved \MinwallaKA, and then their dimension is one. The analysis in four dimensions
can then be carried over to the three-dimensional case almost verbatim, by appropriately
changing the dimensions.  Therefore we again conclude that the conformal manifold close to a given $\cP$ is given by
 \eqn\tmpeqnlabelXXXXXXXXXX{
\cM_c= \{\lambda^i | D^a=0 \} / G
= \{\lambda^i\} / G^{\C}.
}
This  result could find  some applications to the analysis of the theory of M2-branes at the tip of a Calabi-Yau cone.

Although we will not work out the details here, we point out that our analysis can also be extended to $\cN=(2,2)$ superconformal theories in two dimensions.

\newsec{Comparison with the Analysis of Leigh and Strassler}

Having presented our analysis, let us compare it to the classic analysis of exactly marginal deformations presented by Leigh and Strassler \LeighEP.
Their algorithm for determining $\cM_c$ can be summarized as follows:
\item{\bf LS1.}\ List all marginal superpotential couplings $\lambda$ and gauge couplings $\tau$. Take the total number of these complex couplings to be $n$.
\item{\bf LS2.}\ Write down the beta functions for the above couplings. For superpotential couplings, these beta functions are linear combinations of the anomalous dimensions of elementary fields. For gauge couplings, use the NSVZ beta function \NovikovUC, which can also be written in terms of anomalous dimensions.
\item{\bf LS3.}\ Find how many of the beta functions are independent; take the total number of independent conditions to be $k$.
\item{\bf LS4.}\ Find how many phase rotations of $\lambda$ and anomalous shifts of $\tau$ there are.  In practice, one finds by a case-by-case analysis that there are also $k$ of them.
\item{\bf LS5.}\ One concludes that there are at most $n-k$ exactly marginal deformations, because there are $2n$ real coefficients to start with, $k$ real conditions from the beta functions, and $k$ phases to remove.  To determine the precise dimensionality, one must find a point on the manifold where the zeros of the beta functions can be found explicitly.

\medskip

To compare with our method, it is instructive to note that when $\cP$ is the infrared limit of a standard gauge theory, the quantities $D^a$ we used in our discussion are basically the anomalous dimensions of elementary fields.
More explicitly, every elementary chiral superfield $Q$ in a given representation of the gauge group is associated with a current $J=Q^\dagger Q$.  Then the term $\int d^4\theta  Z J$ is the wavefunction renormalization of $Q$, and $D=\mu(\partial/\partial\mu) Z$ is its anomalous dimension. The relations \betafunction\  are related to the standard expressions for the beta functions in terms of anomalous dimensions.

Now let us compare the two methods point by point. We do not differ at LS1.  In \change, we identified the possible form of the renormalization; this step corresponds to LS3. We learned that the independent conditions are associated to broken global symmetry currents.  We used this fact to express all of the beta functions as in \betafunction, thus reproducing LS2. Then, corresponding to LS4, we divided by broken global symmetries. We described why the number of independent conditions and the number of removable phases are generically the same. This fact allowed us to present the conformal manifold as a symplectic quotient.  This step includes LS5, but also provides an algorithm for determining the dimensionality and structure of the manifold at any fixed point.

The main advantage of our technique is that we do not need to refer to the NSVZ beta function. This freedom allows us to address inherently strongly coupled theories. In particular, our analysis applies to deformations of a generic $\cP$ without a known UV Lagrangian description. A related fact is that it is possible to use our approach also in three dimensions. Finally, the identification of the coset structure \hq\ allows us to compute some quantities simply by group-theoretic techniques. For example, to find the dimension of $\cM_c$ we just need to find all the (independent) regular holomorphic symmetry-invariant combinations of $\lambda^i$. Some examples of this procedure are given in the next section.

\newsec{Some Examples}

In this section we consider various examples which highlight the underlying physics of our proof and demonstrate the utility of our result.

\subsec{Wess-Zumino models}

First, consider a theory of a single free chiral superfield $\Phi$.  The free field point has a global $U(1)$ symmetry and conserved current $\Phi^{\dag}\Phi$.  The only marginal operator is $\Phi^3$.  If we deform the theory by the superpotential $\cW =\lambda \Phi^3$, the $U(1)$ symmetry is broken.  Because the symmetry is broken, there is a beta function for $\lambda$, so $\lambda$ must be irrelevant.  In more conventional language, $\lambda \Phi^3$ is not exactly marginal due to wavefunction renormalization which causes $\lambda$ to be irrelevant.  Of course, at free fixed points, the kinetic terms $\Phi^{\dag}\Phi$ are in the same multiplet as a conserved current and so wavefunction renormalization is just one example of current conservation being violated by interactions.

In the language of symplectic quotients, there are no $U(1)$-invariant holomorphic objects that can be constructed out of this single coupling and thus there are no exactly marginal operators. The only marginal operator must by marginally irrelevant.

A slight generalization of this example is also illuminating.  Let $\cP$ be a free theory of $N$ chiral superfields $\Phi^i$.  This theory has $N(N+1)(N+2)/6$ marginal operators $\Phi^i\Phi^j\Phi^k$ and the global symmetry is $U(N)$.  The $N^2$ $D$-term equations are powerful enough to make all the $\cO(N^3)$ marginal operators marginally irrelevant -- none of them is exactly marginal.  To see this, note that the $D$-term equation for $U(1) \subset U(N)$ which rotates all the chiral superfields by a common phase cannot be satisfied.

\subsec{$SU(3)$ with nine flavors}

Next, let us consider one of the original examples of Leigh and Strassler \LeighEP, $N_c =3$ gauge theory with $N_f = 9$ flavors, perturbed by the superpotential
\eqn\tmpeqnlabelXXXXXXXXXXXX{
\cW = \tau \Tr W^\alpha W_\alpha +y [ Q^1 Q^2 Q^3 + Q^4 Q^5 Q^6 + Q^7 Q^8 Q^9 + (Q \leftrightarrow \widetilde{Q})]~.
}

This SCFT is of the form pictured in Figure 1.  From the point of view of Leigh and Strassler, this superpotential was carefully chosen so that there is a single anomalous dimension for all the $Q$'s and the beta function for the gauge coupling $\beta_g \propto \beta_y = 3 \gamma_Q$. Because there is only one independent equation for two couplings, and we can remove one phase, we get a one-dimensional manifold parameterized by $y$.  $\cM_c$ is continuously connected to the free theory of gauge and quark superfields.

Our method allows us to work with a more general superpotential
\eqn\superTN{
\cW = \tau \Tr W^\alpha W_\alpha+\lambda^{a b c} \epsilon_{i j k} Q^{i}_a Q^{j}_b Q^{k}_c + \tilde\lambda^{\tilde a \tilde b \tilde c} \epsilon_{l m n} \widetilde Q^{l}_{\tilde a} \widetilde Q^{m}_{\tilde b} \widetilde Q^{n}_{\tilde c}
}
where $a,\ldots$ and $\tilde a,\ldots$ are flavor indices under the $SU(9) \times SU(9)$ flavor symmetry, and $i,\ldots, n$ are $SU(3)$ gauge indices.  $\lambda^{abc}$ ($\tilde\lambda^{\tilde d \tilde e \tilde f}$) transforms in the three-index antisymmetric representation of the first (second) $SU(9)$.

We can either consider \superTN\ as the deformation of the free theory or a deformation of the SCFT at finite $\tau$ and $y$ of \LeighEP.  In terms of Fig.~1 we expand either around the free point or around $\cP$.

In the first method, we take the theory of free $SU(9)$ gauge fields and $N_f=9$ quarks, and consider the superpotential \superTN.  At the free limit the global symmetry is $U(9)\times U(9)$.  We consider the space of $\{\lambda^{abc},\tilde\lambda^{\tilde a\tilde b\tilde c},\tau\}$ which has complex dimension $2\times 84 + 1 = 169$.  Of the $U(9)\times U(9)$ symmetry,
one linear combination of two $U(1)$'s is anomalously broken by coupling to the gauge field, although in any case both $U(1)$ are broken by non-zero $\lambda$ and $\tilde \lambda$.  We then check that we can turn on $\lambda$ and $\tilde \lambda$, which completely breaks $SU(9) \times SU(9)$ while preserving the  ``D-flatness condition.''  We conclude that all the 162 generators of  $U(9) \times U(9)$ are broken. Therefore, we find $\cM_c$ has complex dimension $169-162=7$. The one-dimensional manifold of Leigh and Strassler is  embedded in $\cM_c$.

We can also use the second method and expand around a point with generic  $\tau$ and $y$.  Here the $SU(9)$ symmetry acting on $Q_a$ is broken to $SU(3)^3$ and similarly for $\tilde Q_{\tilde a}$.  At this point most of the operators of the form \superTN\ are irrelevant.  To see that, note that the currents of the broken global symmetry generators of $U(9)\times U(9) / SU(3)^6$ are paired with chiral operators and become irrelevant.  The remaining marginal couplings are $y$ in \tmpeqnlabelXXXXXXXXXXXX\ and the $SU(3)^3\times SU(3)^3$ breaking marginal operators are $\lambda^{abc}Q_{a}Q_b Q_c$ with $a=1,2,3$, $b=4,5,6$ and $c=7,8,9$, and similarly for $\tilde\lambda^{\tilde a\tilde b\tilde c}$.  These provide 54 additional marginal operators.  We can break all of $SU(3)^3\times SU(3)^3$ preserving  the ``D-flatness condition.''  Therefore, we have $54- 48 = 6$ additional exactly marginal operators and $\cM_c$ has complex dimension $6+1=7$.

This  example should highlight the major advantage of our methods over the methods of \LeighEP.  Had we followed Leigh and Strassler, we would have been forced to analyze all 169 beta functions to determine that only 162 were linearly independent. From analyzing the beta functions directly, one might be led to believe that what makes the manifold possible is the high degree of symmetry of the theory.  In some sense, the opposite is true.  The global symmetry of our original fixed point is the only thing that prevents marginal operators from being exactly marginal.  Furthermore, at generic points on the conformal manifold, there is no global symmetry and the superpotential does not have any symmetric form.

\subsec{SQCD in the conformal window}

Our next example is the infrared limit of the $\cN=1$ $SU(N)$ theory with $N_f$ quarks $Q,\widetilde Q$,  when $3N_c/2 < N_f < 3N_c$.  This is  known to be an interacting superconformal theory~\SeibergPQ, which we take as  $\cP$.  The situation is as in Fig.~2.

To study its deformation, we need to list marginal chiral operators.  This can be done by studying what happens to the operators of the UV Lagrangian description.  For general $N_f$, there are no chiral marginal operators; $\Tr W^\alpha W_\alpha$ is not a chiral primary operator in $\cP$.  There are no marginal gauge-invariant chiral primaries constructed out of $Q$ and $\tilde Q$ either.

However, in the special case $N_f=2N_c$, the operator $\cO^{ij}_{\tilde k\tilde l}=(Q^i\tilde Q_{\tilde k})(Q^j\tilde Q_{\tilde l})$ is marginal at $\cP$.  We can therefore deform it by
\eqn\confwi{\cW=\lambda_{ij}^{\tilde k \tilde l} \cO^{ij}_{\tilde k\tilde l}~.}  The couplings $\lambda_{ij}^{\tilde k \tilde l}$ transform under the flavor symmetry $SU(N_f)^2$; the pairs $i,j$ and $\tilde k, \tilde l$ are either both symmetric or both anti-symmetric.  Therefore there are in total $[N_f(N_f+1)/2]^2+[N_f(N_f-1)/2]^2=N_f^2(N_f^2+1)/2$ marginal operators.  (Note that the baryon number $U(1)$ symmetry is not broken by these deformations.)

The analysis of the D-flat conditions is best done, as in the previous example, in two steps.  Consider first the special choice \LeighEP\ $\lambda_{ij}^{\tilde k \tilde l}\propto \delta^{\tilde k}_i\delta^{\tilde l}_j$.  It is easy to see that this is a D-flat direction which breaks the symmetry to the diagonal $SU(N_f)$.   $N_f^2-1$ of the operators in \confwi\ are lifted and are no longer marginal.  The remaining marginal operators can be classified according to the remaining $SU(N_f)$ symmetry.  Then it is again easy to find D-flat directions which completely break this symmetry.

We conclude that the quotient by complexified $SU(N_f)^2$ flavor symmetry  removes $2N_f^2-2$ directions leaving many (order $N_f^4/2$) exactly marginal directions.

\subsec{Conifold theory}

As another example, let us consider the conifold theory \KlebanovHH.  The situation is again as in Fig.~2.  We start from an $SU(N) \times SU(N)$ gauge theory  with two vector-like pairs of bifundamentals $A_{a}, B_{\dot{a}}$, where $a, \dot{a}$ run over the $SU(2) \times SU(2)$ flavor indices.  We take the gauge couplings $g_1,g_2$ of the two $SU(N)$ groups to be the same and add to this theory the superpotential \KlebanovHH\
\eqn\tmpeqnlabelXXXXXXXXXXXXX{
\cW _{\rm KW} = \lambda_{\rm KW} \epsilon^{a b} \epsilon^{\dot{a} \dot{b}} \Tr (A_{a} B_{\dot{a}} A_{b} B_{\dot{b}})~.
}

This theory flows to a conformal fixed point in the IR with an unbroken $SU(2) \times SU(2) \times U(1)$ global symmetry, which we take to be the reference point $\cP$.

Marginal chiral operators which preserve $SU(2)\times SU(2)$
are $\cW_{\rm KW}$ and $\Tr W^{\alpha}_{(1)}W_{ \alpha (1)}-\Tr W^{\alpha}_{(2)}W_{\alpha (2)}$.  The other combination of gauge kinetic terms is not chiral primary because of the anomaly. Therefore the part of $\cM_c$ which does not break $SU(2)\times SU(2)$ is of complex dimension two.

We also have marginal operators $\cO_{(ab),(\dot a\dot b)}=\Tr( A_{a} B_{\dot{a}} A_{b} B_{\dot{b}})$ which break $SU(2)\times SU(2)$.  The trace requires that the coupling is symmetric separately in the dotted and undotted indices (denoted by the brackets).  This coupling transforms in the $({\bf 3}, {\bf 3})$ representation of the $SU(2) \times SU(2)$ but is invariant under $U(1)$ baryon symmetry.  We can deform  $\cP$ by adding these to the superpotential \BenvenutiWI\
\eqn\tmpeqnlabelXXXXXXXXXXXXXX{
\cW= \cW_{\rm KW} + \lambda^{(ab), (\dot{a} \dot{b})} \cO_{(ab),(\dot a\dot b)}~.
}
For generic values of $\lambda$, the unbroken global continuous symmetry is $U(1)$.

Using our general result, this case is also easy to understand.  We have 11 marginal operators.  These deformations completely break $SU(2) \times SU(2)$.
Thus we have $\cM_c$ which is of complex dimension $11-6 = 5 $.  This same result was found in \BenvenutiWI\ using the Leigh-Strassler method.

\subsec{$E_6$ theory}

Now let us consider an example whose analysis is impossible using the Leigh-Strassler method.  Minahan and Nemeschansky studied an interesting $\cN=2$ superconformal field theory with $E_6$ symmetry \MinahanFG.  (A Lagrangian description of this theory was recently found in \ArgyresCN.)  This theory has one dimension-three  chiral primary  $u$  parameterizing the Coulomb branch, and $78$ dimension-two chiral primaries  $\mathbb{X}$ transforming in the adjoint of $E_6$ parameterizing the Higgs branch.  Therefore the operator $u$ is marginal. As we now describe, it is however not exactly marginal.

The reason is as follows. The $\cN=2$ superconformal algebra has
R-symmetry $SU(2)_R\times U(1)_R$. Denote the generators of these groups by $I^{a=1,2,3}$ and $R_{\cN=2}$, respectively. The generator of the $U(1)_R$ symmetry of the $\cN=1$ subalgebra is $R_{\cN=2}/3 + 4I^3/3$, and another linear combination $J=R_{\cN=2}-2I_3$ becomes a non-R flavor  symmetry from the point of view of $\cN=1$ theory. $u$ is charged under this $J$.  Therefore $u$ becomes marginally irrelevant.

\newsec{Conclusions}

In this note, we found that the space of exactly marginal operators of an $\cN=1$ SCFT is given by the quotient of the space of marginal couplings by the complexified global symmetry $G^{\C}$. Additionally, we described a few examples where this description gives a streamlined way to count the number of exactly marginal couplings. We now conclude by considering a few possible future directions.

First, we note that our method provides a convenient way to study the conformal manifold of any four-dimensional $\cN=1$ SCFTs, and opens the door to understanding many new SCFTs.  It can also be extended to three-dimensional $\cN=2$ and two-dimensional $\cN=(2,2)$ superconformal theories.

Second, it might be interesting to see if we can show that the renormalization group is a gradient flow to all orders in a particular renormalization scheme. That the RG flow is a gradient flow in two dimensions has been known  for a long time. There have been efforts to extend this statement to four-dimensional $\cN=1$ theories, e.g.~\refs{\KutasovUX,\BarnesJJ}. It is also known that under AdS/CFT, the holographic renormalization group is driven by a gradient flow, as in e.g.~\FreedmanGP.  In standard worldsheet string theory and in AdS/CFT there is a clear dictionary between the properties of the CFT and the target space.  The gradient flow is determined by an action in the space of coupling constants of the CFT.  Our analysis, which is more general, also points in the direction of such an effective action in the space of coupling constants.  Around \tmpeqnlabelXXXXXXX\ we have discussed this effective action as a function of the couplings $\lambda^i$.  We can view this action as arising from integrating out the auxiliary $D$-terms of the gauge multiplets $Z^a$.  Alternatively, we can write an action which depends on $Z^a$.  It includes a K\"ahler potential which depends on $\lambda^i$, $\bar \lambda^{\bar i}$, and $Z^a$, and a term $\int d^2 \theta \gamma_{ab} W_\alpha^a(Z) W^{b\alpha}(Z)$, where $W_\alpha^a(Z)$ is the field strength of the gauge field $Z$.

It would be nice to understand this action in more detail and to see where else such an action is present.

Finally, one might hope to find the Zamolodchikov metric on the conformal manifold, instead of just the structure as a complex manifold. This metric is known to be K\"ahler \AsninXX, but there might also be additional structure, as was the case in two-dimensional $\cN=(2,2)$ supersymmetry \refs{\CecottiZX\CecottiQN\DixonFJ{--}\StromingerPD} and as was recently argued in four-dimensional $\cN=2$ theories \PapadodimasEU.

We hope to come back to these matters in the future.

\bigskip

\centerline{\bf Acknowledgements}
We would like to thank O.~Aharony, K.~Intriligator, D.~Jafferis, J.~Maldacena, and M.~Strassler for helpful discussions.
DG, NS, and BW are supported in part by DOE grant DE-FG02-90ER40542. ZK and YT are supported in part by NSF grant PHY-0503584. YT is additionally supported by the Marvin L. Goldberger membership at the Institute for Advanced Study. BW is additionally supported by the Frank and Peggy Taplin Membership at the Institute for Advanced Study.

\appendix{A}
{Classification of Supersymmetric Deformations}
In this appendix we classify possible terms which can be added to the Lagrangian of a superconformal field theory.
We will find that the only possible deformations are those of the superpotential and the K\"ahler potential, as is usually assumed.
Furthermore, K\"ahler deformations turn out to be always irrelevant.

As a preliminary we first review some properties of the representations of the $d=4$, $\cN=1$ superconformal algebra following \refs{\FlatoTE,\DobrevQV}\foot{Also see \refs{\MinwallaKA,\OsbornQU} and  Appendix B of \FreedmanGP.}.
A primary operator is characterized by its left and right spin $j$, $\tilde\jmath$
with indices $\alpha$ and $\dot\alpha$ respectively,
and by its $U(1)_R$ charge $r$ and the dimension $d$.
Positivity of the norm of its descendants leads to unitarity bounds. Important ones are as follows:
\eqnn\stateone
\eqnn\statetwo
\eqnn\statethree
\eqnn\statefour
\tabskip=0em plus 1fil
\halign to\hsize{  for $#$,   &  $#$    & leads to\quad\quad  $#$  \hfil & \hfill # \tabskip=0pt\cr
\tilde\jmath =0 & [Q^\dagger _{\dot\alpha} , \cO_{j,\tilde\jmath=0,r }\}|_{\half}
&  d  \ge\phantom{-} {3\over 2} r \, , &\stateone \cr
j =0 & [Q _\alpha ,\cO_{j=0,\tilde\jmath,r}\} |_{  \half}
& d \ge - {3\over 2} r \, , & \statetwo \cr
\tilde\jmath > 0 & [Q^\dagger _{\dot\alpha}, \cO_{j,\tilde\jmath,r}\} |_{\tilde\jmath - \half}
& d \ge \phantom{-}{3\over 2} r + 2 + 2\tilde\jmath \, , &\statethree\cr
j>0 &  [Q _\alpha,\cO_{j,\tilde\jmath,r}\}|_{ j - \half}
& d  \ge -{3\over 2} r + 2 + 2  j \, &\statefour\cr
}

Here  $[...\}$ denotes either a commutator or an anticommutator and
$...|_{\tilde\jmath +\half}$ denotes that the indices are contracted to this spin.  The operators $\cO_{j,\tilde\jmath,r}$ are only assumed to be primary and are otherwise generic.

If $\tilde \jmath=0$ and the operator in  \stateone\  does not vanish,
we should also study
\eqn\statethreep{
  [Q^\dagger ,[ Q^\dagger, \cO_{j,\tilde\jmath=0,r}\}\}
}
 whose positivity leads to
\eqn\conconda{
d \ge {3\over 2} r + 2 ~.
}
Similarly, if $j=0$ and  the operator  \statetwo\  does not vanish,
we should also study
\eqn\statefourp{
 [Q, [Q,\cO_{j=0,\tilde\jmath,r}\}\}
}
 which leads to
\eqn\concondana{
d \ge -{3\over 2} r + 2.
}

Each inequality is saturated when the corresponding descendant vanishes; the descendant is a null vector.
In particular, if \stateone\  vanishes, the operator is called \emph{chiral}.
If \statetwo\  vanishes, the operator is called \emph{antichiral}.
If \statethree\  or \statethreep\  vanishes the operator is called \emph{left semi-conserved}
and
if \statefour\  or \statefourp\  vanishes it is called \emph{right semi-conserved}.
Operators which are both left and right semi-conserved are called \emph{conserved}.
They satisfy $d=2+j + \tilde\jmath$ and $r = {2\over 3} (j-\tilde\jmath)$.

The inequalities above guarantee that all other descendants, e.g.~
\eqn\stateonep{
[Q^\dagger _{\dot\alpha} , \cO_{j,\tilde\jmath,r }\}
_{\tilde\jmath + \half} \qquad {\rm for} \qquad  \tilde\jmath>0
}
do not lead to additional constraints, and in particular, they cannot vanish.

Now we examine the possible supersymmetric operators which can deform the Lagrangian.  Imposing Lorentz invariance, the candidate operators are
\eqn\dummy{\CO_0}
\eqn\dummyA{\{Q,[Q, \CO_1]\}}
\eqn\dummyB{\{Q^\dagger,[Q^\dagger,\{Q,[Q, \CO_2]\}]\}}
\eqn\QU{\{Q^\alpha, U_\alpha\}}
\eqn\QQY{Q^{\dagger \dot\alpha},[Q^\alpha, Y_{\alpha\dot\alpha}]\}}
\eqn\QQQZ{\{Q^\dagger,[Q^\dagger,\{Q^\alpha, Z_\alpha\}]\}}
and their complex conjugates.
Without loss of generality we can take the operators $\cO_0$,
$\cO_1$, $\cO_2$, $U_\alpha$, $Y_{\alpha\dot\alpha}$ and $Z_\alpha$
to be primary operators.
Note that the order of the supersymmetry generators in the multiple commutators does not matter
-- changing it adds to the Lagrangian a total derivative.

Let us impose supersymmetry invariance.  It is important that the only facts we can use are the null vectors mentioned above; i.e.~the vanishing of some operators of the form~\stateone\ -- \statefourp.

Supersymmetry invariance of $\cO_0$ requires that it be annihilated both by $Q$ and by $ Q^\dagger$, and hence it must be the identity operator.
This changes the vacuum energy but does not affect the dynamics of the theory.  Hence this case can be ignored.

Supersymmetry invariance of $\cO_1$ requires that it be chiral and this is the standard superpotential deformation.

The operator derived from $\cO_2$ is not constrained by supersymmetry;
this is the standard K\"ahler deformation.
However, if we want it to be nontrivial, none of the null vectors discussed above are present.  In particular, it cannot be chiral and it cannot be semi-conserved.
This means that $d >{3\over 2}|r| + 2$ and hence it is an irrelevant operator.

For $U_\alpha$, the invariance of \QU\  under $Q$ means that
\eqn\tmpeqnlabelXXXXXXXXXXX{
[ Q_\beta, \{ Q^\alpha, U_\alpha\}]=\partial_{\mu}(\cdots).
}
However, none of the null vector conditions above leads to this condition with a non-zero total derivative.
Since $U_\alpha$ has $j=\half$, we must use \statefour\ ,
but this means that the operator \QU\  itself vanishes.
Hence such an operator should not be counted among the deformations of the Lagrangian.

Similarly, for $Y_{\alpha\dot\alpha}$ and $Z_{\alpha}$,
invariance of \QQY\  or \QQQZ\  under $Q$ requires the use of
\statefour\ , which makes \QQY\  and \QQQZ\  themselves to vanish.
Therefore, operators $U_\alpha$, $Y_{\alpha\dot\alpha}$ and $Z_\alpha $
do not lead to independent deformations.

We conclude that $\{Q^\alpha,[Q_\alpha, \cO_1 ]\}$ is the only possible marginal or relevant deformation of a unitary SCFT.

 \listrefs
\bye